\documentclass[preprint,10pt]{elsarticle}

\journal{Nuclear Physics A}

\usepackage{amssymb}

\usepackage{lineno}
\usepackage{float}
\usepackage{placeins}

\usepackage[figuresright]{rotating}
\usepackage{subcaption}

\def\pt{$p_{\mathrm{T}}$}
\def\sqs{$\sqrt{s}$~}
\def\sqsnn{$\sqrt{s_{\rm{NN}}}$~}

\def\raa{$R_{\mathrm{AA}}$~}
\def\v{$\nu_{2}$~}

\begin{document}

\begin{frontmatter}



\title{Heavy-flavour decay lepton measurements in pp, p--Pb, and Pb--Pb collisions with ALICE at the LHC}

\author{S. LaPointe on behalf of The ALICE Collaboration}
\address{Istituto Nazionale di Fisica Nucleare, Via P. Giuria 1, 10125 Torino}

\begin{abstract}

We present the measurements of electrons and muons from the semi-leptonic decays of heavy-flavour hadrons measured in the central and forward rapidity regions with ALICE in pp (\sqs = 2.76 TeV and 7 TeV), Pb--Pb (\sqsnn = 2.76 TeV), and p--Pb (\sqsnn = 5.02 TeV), collisions at the LHC. The \pt-differential production cross section in pp collisions, the elliptic flow in Pb--Pb collisions, and the nuclear modification factor in Pb--Pb and p--Pb collisions are shown. The results are compared to theoretical predictions.

\end{abstract}

\end{frontmatter}


\section{Introduction}
\label{intro}

The overall objective of ALICE is to study the QCD matter that forms from the collisions of Pb ions in the CERN LHC \cite{alice}. Heavy quarks, i.e. charm and beauty, are considered to be excellent probes of such matter as they are predominantly produced via hard parton scatterings, in the initial phase of the collision.  They are expected to lose energy through elastic scattering and radiative processes. 

An experimental observable sensitive to the energy loss is the nuclear modification factor ($R_{\mathrm{AA}}$), which is the ratio of particle yields in heavy-ion collisions to those in pp collisions, scaled by the number of binary collisions. It is expected to be equal to unity in the absence of initial and final state medium effects. Another experimental observable of interest is the elliptic flow ($\nu_2$), which is the azimuthal anisotropy in momentum space that arises from an ``almond'' shaped overlap region of a non-central heavy-ion collision expanding under anisotropic pressure gradients. $\nu_2$ is used to quantify in-medium collectivity. For an interacting, collectively expanding medium created in a non-central collision \v should be positive. Additionally, at high \pt, \v is sensitive to the path--length dependence of in-medium energy loss. 

The measurements in pp collisions provide a baseline for the Pb--Pb and p--Pb collisions while also offering key tests of pQCD predictions of heavy quark production. The p--Pb studies are crucial for the interpretation of the Pb--Pb results, as they allow us to disentangle cold nuclear matter effects from those effects related to the formation of QCD matter. 

In the following the ALICE measurements of electrons and muons originating from the semi-leptonic decays of heavy-flavour hadrons in pp, Pb--Pb, and p--Pb collisions are presented. The measured \pt-differential production cross sections, nuclear modification factor, and elliptic flow are shown and compared to the respective theoretical predictions.
%
%

\section{Heavy-flavour measurements with electrons and muons in ALICE}
\label{hfALICE}

The heavy-flavour particle detection performance of ALICE is detailed in \cite{ppr2}. For the results presented here, the detector components utilized are located in the central rapidity ($|\eta|$ $<$ 0.9) and forward rapidity ($-$4 $<$ $\eta$ $<$ $-$2.5) regions. The central region consists of the Inner Tracking System, surrounded by the Time Projection Chamber (TPC). Electrons are identified with TPC and Time of Flight (TOF) and at higher momentum using the ElectroMagnetic Calorimeter (EMCal) and the Transition Radiation Detector. In the forward region muon tracking and identification is provided by the muon spectrometer. 

Open heavy-flavour hadrons are measured via their semi-leptonic decay channels in the central rapidity region with electrons and in the forward region with muons. Concerning the electrons, the background is dominated by Dalitz decays and photon conversions in detector material. These sources were estimated using two different techniques, one involving a cocktail of electron background sources and the other utilizing an e$^+$e$^-$ invariant mass method \cite{alice_hfe7}. The dominant background sources for the muon measurement are from the decays of pions and kaons. They are estimated in pp collisions using a MC simulation, while in Pb--Pb collisions they are extrapolated from the measured $\pi$, K yields at central rapidity. Additionally, ALICE is capable of identifying the electrons from the decay of beauty hadrons by exploiting the relatively long lifetime of B mesons for a track displacement-based selection and the different decay kinematics of the D and B mesons in a $\Delta\varphi$ correlation-based approach.

%
%

\section{Measurements in pp collisions}
\label{pp}

\begin{figure}[!htb]
\begin{subfigure}[b]{.5\textwidth}
\includegraphics[width=1\linewidth]{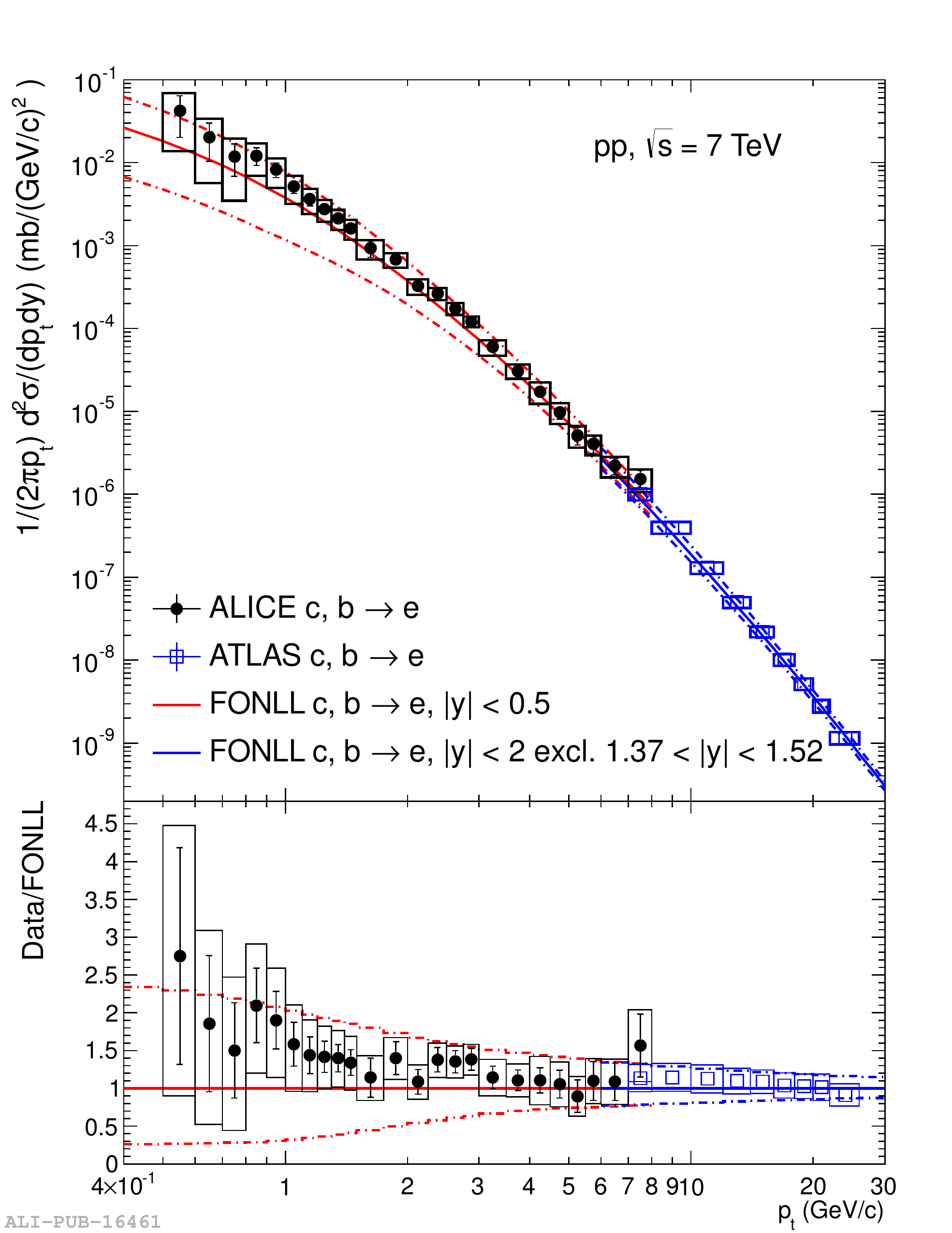}
\end{subfigure}%
\begin{subfigure}[b]{.5\textwidth}
  \includegraphics[width=1\linewidth]{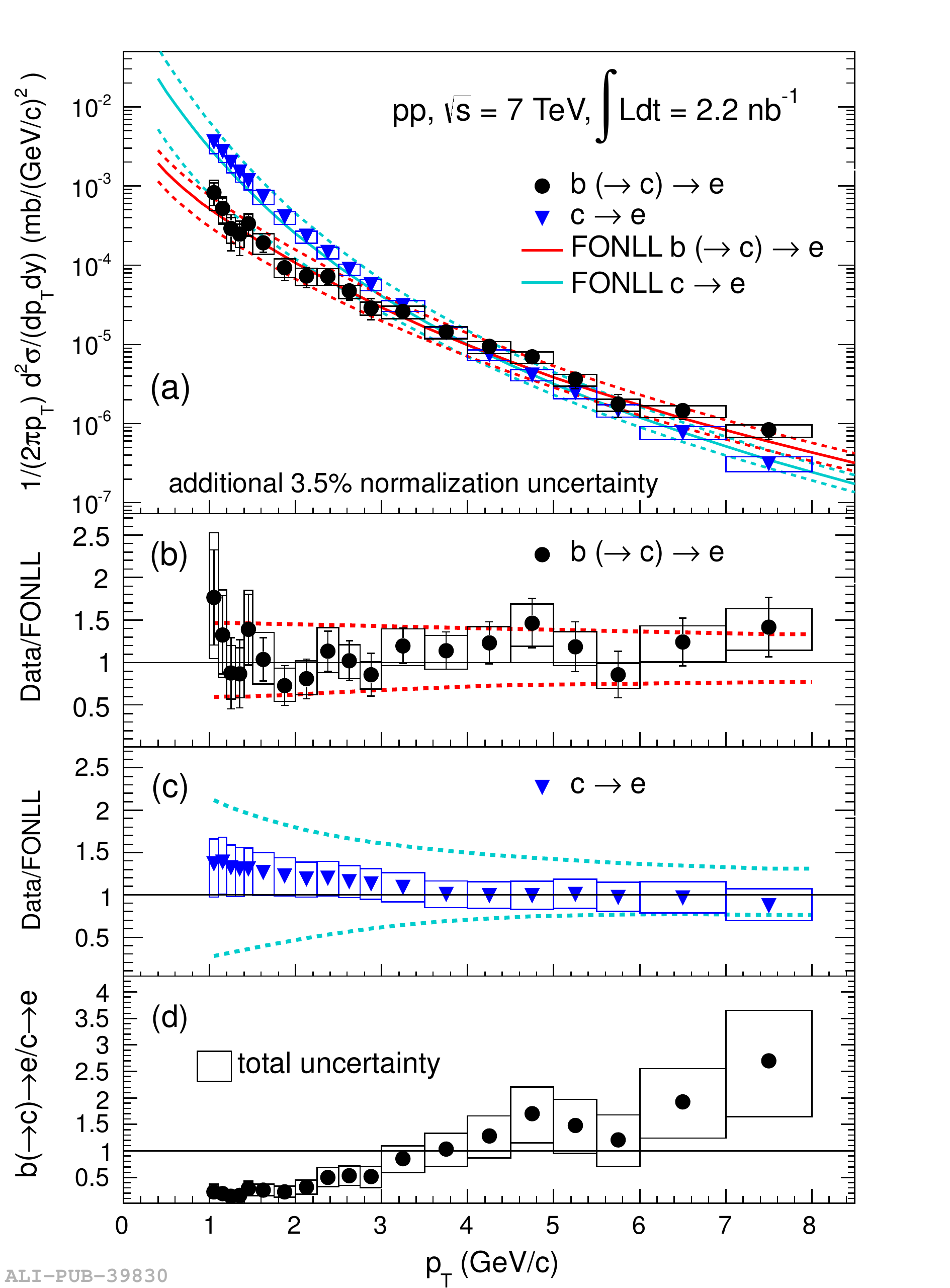}
\end{subfigure}
\caption{(Color online) Left: \pt-differential production cross section of electrons from heavy-flavour hadron decays from ALICE in $|y| < $ 0.5 and ATLAS in $|y| < $ 2, excluding 1.37 $< |y| < $ 1.52 \cite{atlas}. The solid (dashed) lines show the corresponding FONLL predictions (uncertainties). The lower panel shows the ratio of the data to the prediction. Right: (a) \pt-differential cross section of electrons from beauty  and charm hadron decays measured in pp collisions at \sqs = 7 TeV.  The solid (dashed) lines indicate the corresponding FONLL predictions (uncertainties). Ratios of the data and the FONLL calculations are shown in (b) and (c) for electrons from beauty and charm hadron decays, respectively, where the dashed lines indicate the FONLL uncertainties. (d) Measured ratio of electrons from beauty and charm hadron decays with error boxes depicting the total uncertainty \cite{beauty7pp}.}
\label{fig:leptonspp}       
\end{figure}

The \pt-differential production cross section of electrons from heavy-flavour hadron decays in pp collisions at \sqs = 7 TeV at central rapidity ($|y| < $ 0.5) and in the \pt~ range 0.5--8 GeV/$c$ \cite{alice_hfe7} is shown in Fig \ref{fig:leptonspp} (left). The measurement is compared to the complementary result of the ATLAS experiment \cite{atlas}, measured in the rapidity interval $|y| < $2 (excluding 1.37 $< |y| < $ 1.52), which spans a \pt~ range 7--26 GeV/$c$. The corresponding pQCD FONLL predictions \cite{fonll3} in the respective rapidity intervals are also shown, along with the ratio of data to FONLL. Both data samples, within the experimental and theoretical uncertainties, are well described by the predictions. ALICE has also measured muons from the decays of heavy-flavour hadrons in pp collisions at \sqs = 7 TeV at forward rapidity ($4 < y < 2.5$)  \cite{alice_hfm}. 

Additional heavy-flavour measurements in pp collisions include electrons from beauty hadron decays at $\sqrt{s}$ = 7 TeV. They are selected based on the track impact parameter, which is the distance of closest approach of the track to the interaction vertex \cite{beauty7pp}. The \pt-differential production cross section measured is shown in Fig. \ref{fig:leptonspp} (right), along with the production cross section of electrons from charm hadron decays, which was calculated using the charm hadrons measured by ALICE \cite{charmpp} and PYTHIA decay kinematics. The results are compared to the FONLL predictions \cite{fonll3} and agree within the theoretical and experimental uncertainties. 

\begin{figure}
 \centering
\includegraphics[width=0.4\linewidth]{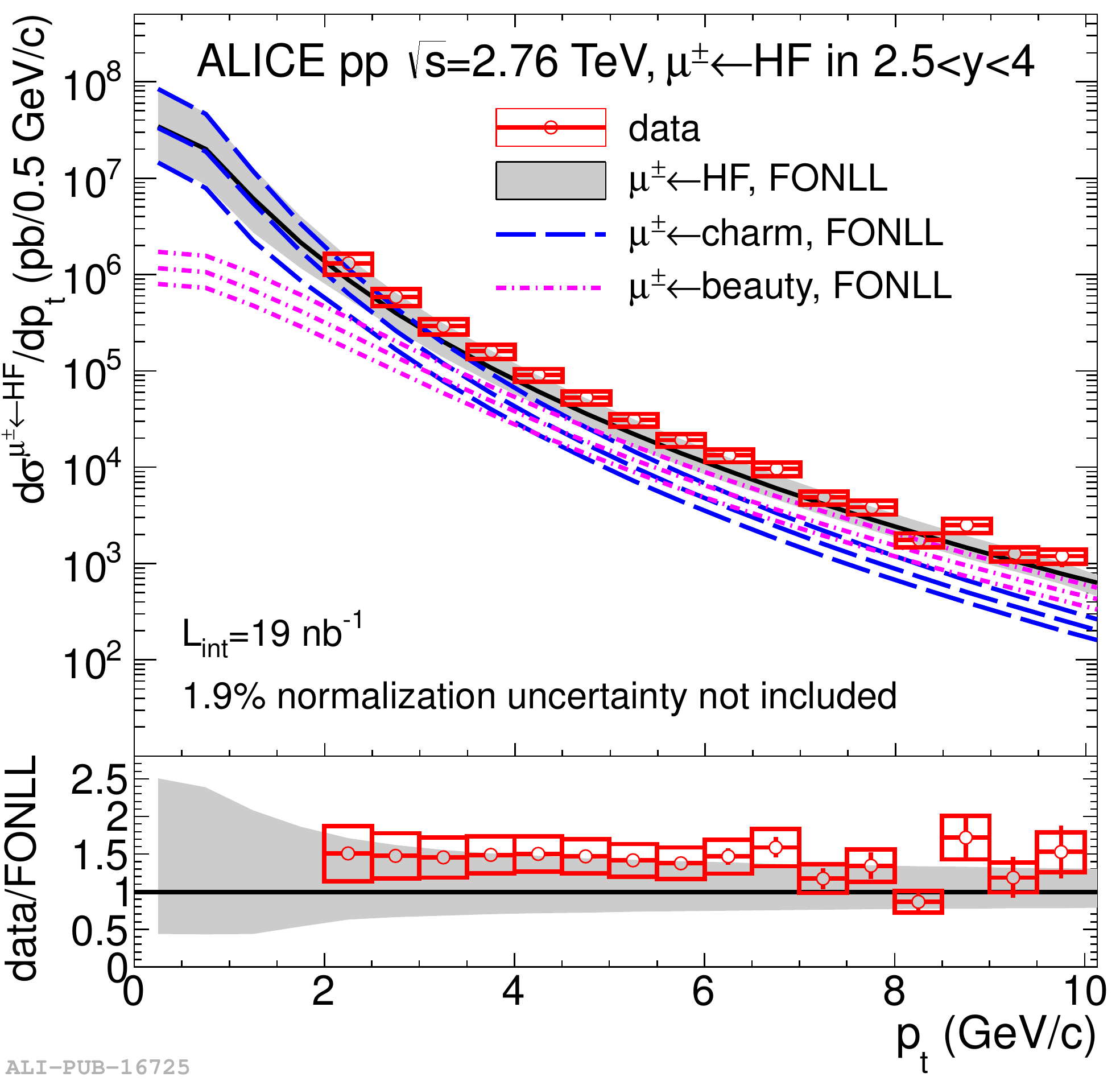}

\caption{(Color online) \pt-differential cross section of muons from heavy-flavour hadron decays in 2.5 $ < y < $ 4 in pp collisions at \sqs = 2.76 TeV. The data are compared to FONLL predictions, with the lower panel showing the ratio between data and the calculation.}
\label{fig:lep276}       
\end{figure}

Heavy-flavour decay leptons have been measured in pp collisions at \sqs = 2.76 TeV, the reference energy for Pb--Pb collisions. The running time for this data sample was short, and consequently the uncertainties of the electron spectrum from the \sqs = 7 TeV sample are significantly lower compared to those at \sqs = 2.76 TeV. Hence the spectrum at \sqs = 7 TeV was scaled down to 2.76 TeV using the FONLL prediction and used as the pp reference for Pb--Pb measurements. The resulting scaled spectrum was compared to the measured spectrum at \sqs = 2.76 TeV and it was found to be in agreement, validating the use of the scaling method \cite{fonllscaling}. The \pt-differential production cross section of muons from heavy-flavour hadron decays in pp collisions at \sqs = 2.76 TeV, measured in the \pt~ range 2--10 GeV/$c$ and 2.5 $ < y < $ 4 is shown in Fig. \ref{fig:lep276}. The measurement is described by the corresponding FONLL predictions \cite{fonll3} and had a statistical precision better than that of the Pb--Pb measurement. Therefore, it is directly used as the pp reference.

%
%

\section{Measurements in Pb--Pb collisions}
\label{PbPb}

The analysis strategy implemented for Pb--Pb data resembles that used for pp data. In the following, the resulting nuclear modification factor and the elliptic flow of electrons and muons from heavy-flavour hadron decays are presented.

\begin{figure}
\begin{subfigure}[b]{.5\textwidth}
\includegraphics[width=1\linewidth]{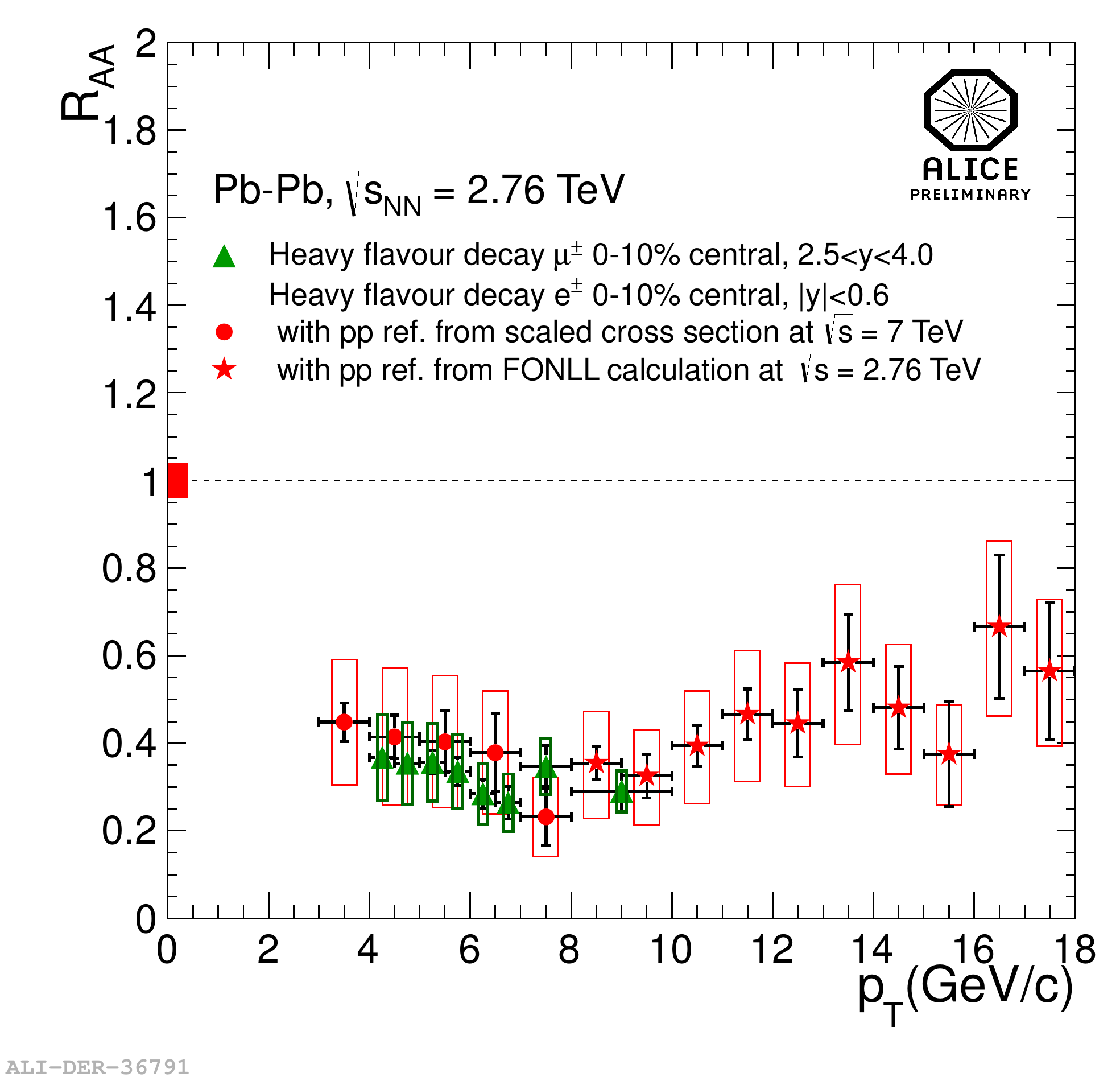}
\end{subfigure}%
\begin{subfigure}[b]{.5\textwidth}
  \includegraphics[width=1\linewidth]{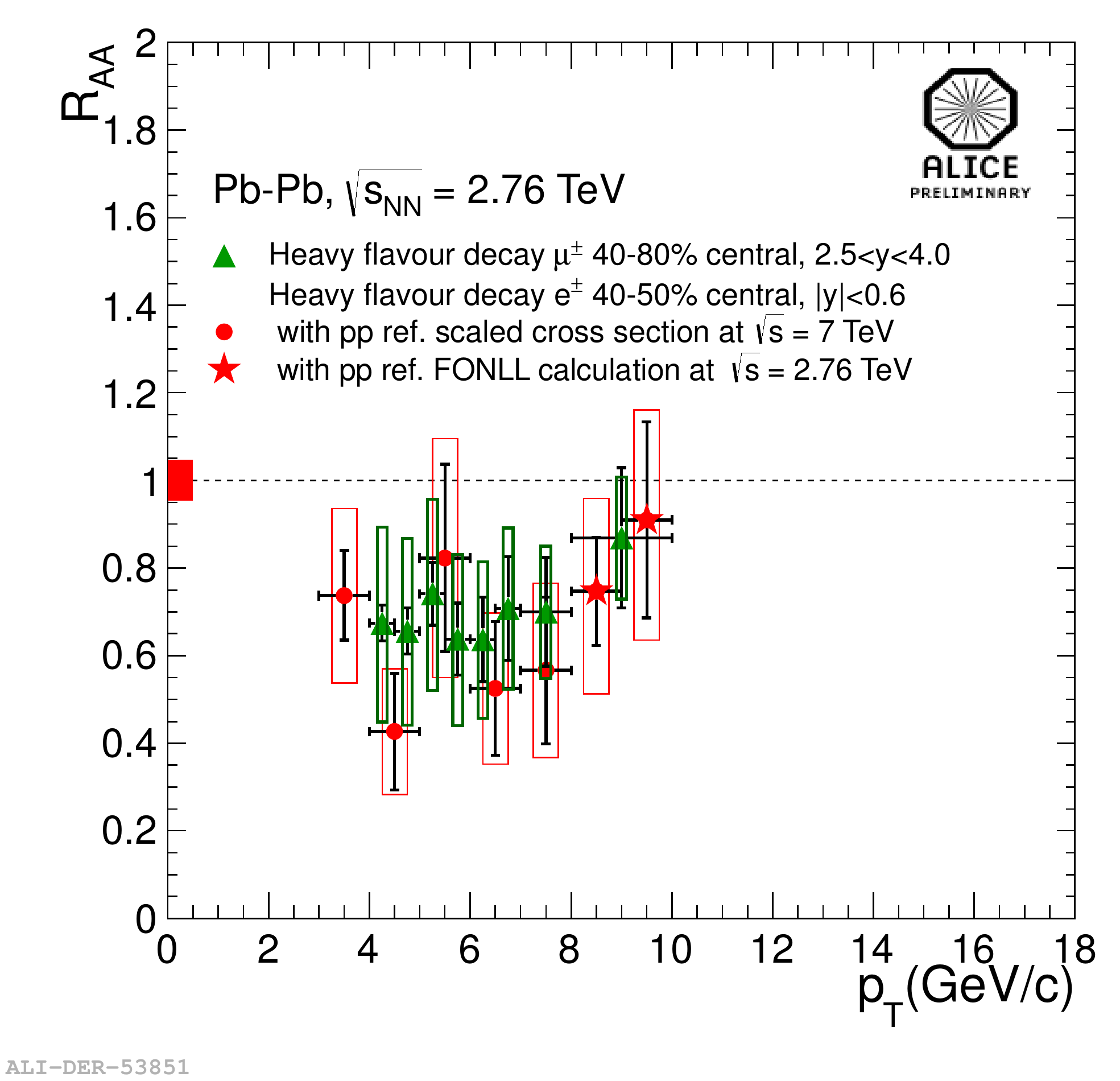}
\end{subfigure}
\caption{(Color online) Left: Nuclear modification factor (R$_{\mathrm{AA}}$) of electrons (circles and stars, depending on the pp reference) and muons \cite{alice_hfm7} (triangles) from heavy-flavour hadron decays in Pb--Pb collisions at \sqsnn = 2.76 TeV for the centrality class 0--10$\%$. Right: The R$_{\mathrm{AA}}$ of electrons measured in the 40--80$\%$ and muons in the 40--50$\%$ centrality classes. For both panels, the muons are measured in the forward rapidity region (2.5 $ < y < $ 4), while the electrons are observed at central rapidity ($|y| < $ 0.6).}
\label{fig:leptonRaa}       
\end{figure}

The nuclear modification factor (\raa) is defined as:

\begin{equation}
	R_{\mathrm{AA}} (p_\mathrm{T}) = \frac{1}{\langle T_{\mathrm{AA}} \rangle} \frac{\mathrm{d}N_{\mathrm{AA}}/\mathrm{d}p_\mathrm{T}}{\mathrm{d}\sigma_{\mathrm{pp}}/\mathrm{d}p_\mathrm{T}}
\end{equation}

\noindent
where d$N_{\mathrm{AA}}$/d$p_\mathrm{T}$ is the yield in Pb--Pb collisions, d$\sigma_{\mathrm{pp}}$/d$p_{\mathrm{T}}$ is the differential cross section in pp collisions, and $\langle T_{\mathrm{AA}} \rangle $ is the average nuclear overlap function, calculated using the Glauber model \cite{aliceGlauber}. In Pb--Pb collisions the \pt-differential yield of heavy-flavour decay electrons is measured at central rapidity in the range 3 $<$ $p_{\mathrm{T}}$ $<$ 18 GeV/$c$. As the measured cross section extends past 8 GeV/$c$ (the upper limit of the measurement in pp at \sqs = 7 TeV), the FONLL calculation was used as the reference for \pt~ $> $ 8 GeV/$c$. The muons are measured in the forward region in the range 4 $<$ $p_{\mathrm{T}}$ $<$ 10 GeV/$c$. Fig. \ref{fig:leptonRaa} (left) shows the measured \raa of the heavy-flavour decay electrons and muons for the centrality class 0--10$\%$. Both the muons and electrons reach a suppression factor 3-4 for \pt~ $> $ 5 GeV/$c$. It is important to note that for \pt~ $>$ 5 GeV/$c$ FONLL calculations predict that the leptons from B meson decays start to dominate the \pt~ spectrum, thus hinting at a significant in-medium energy loss for beauty quarks. Also shown in Fig. \ref{fig:leptonRaa} (right) is the resulting \raa measured in peripheral collisions, of muons in the 40--80$\%$ and electrons in the 40--50$\%$ centrality classes. Compared to the most central collisions, the suppression factor is smaller and indicates an approach toward unity with decreasing collision centrality. This is expected, as the medium formed in peripheral collisions, relative to central, should not be as dense, resulting in less modification of the \pt~ spectrum.

\begin{figure}
\begin{subfigure}[b]{.5\textwidth}
\includegraphics[width=1\linewidth]{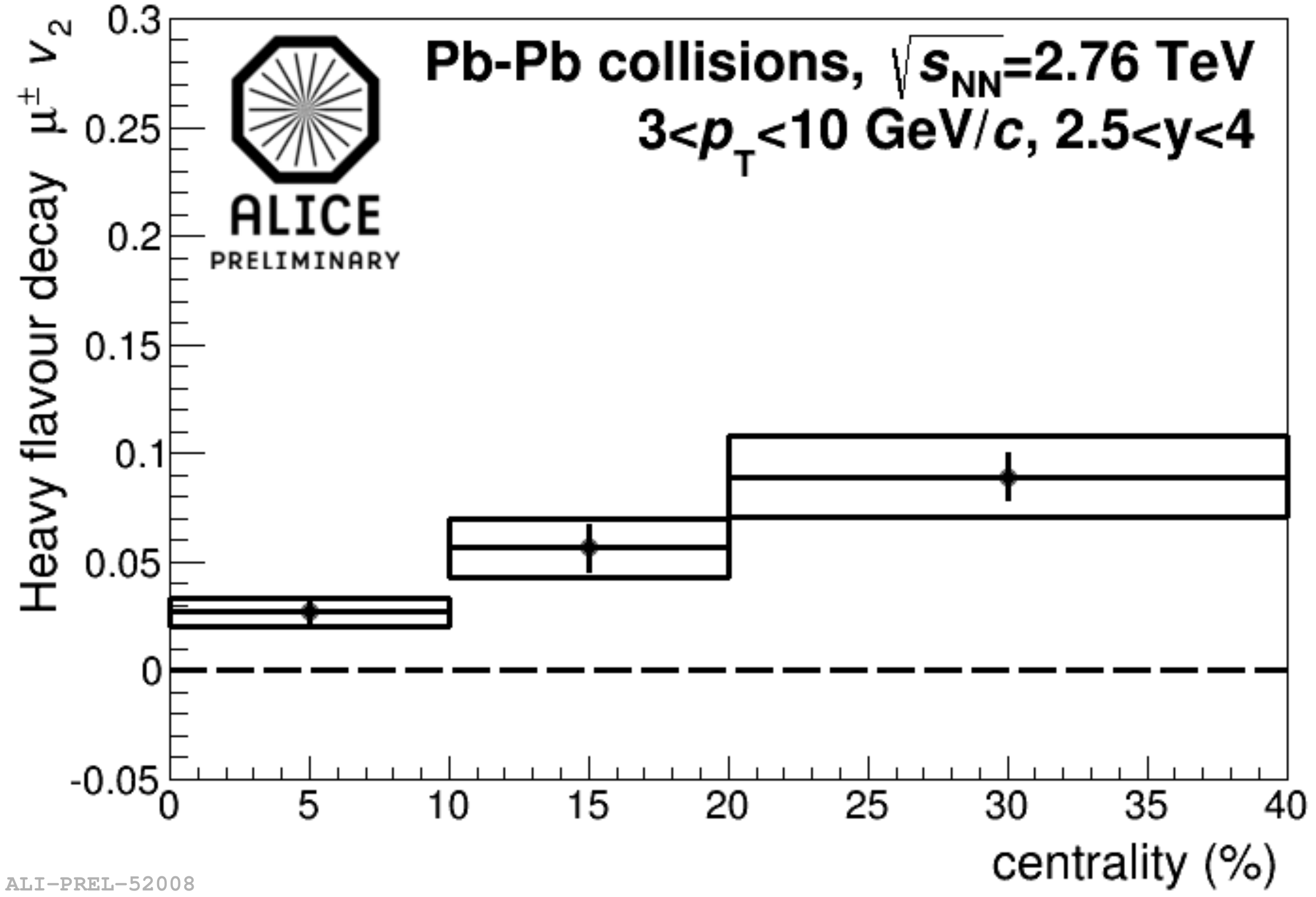}

\end{subfigure}%
\begin{subfigure}[b]{.5\textwidth}
  \includegraphics[width=1\linewidth]{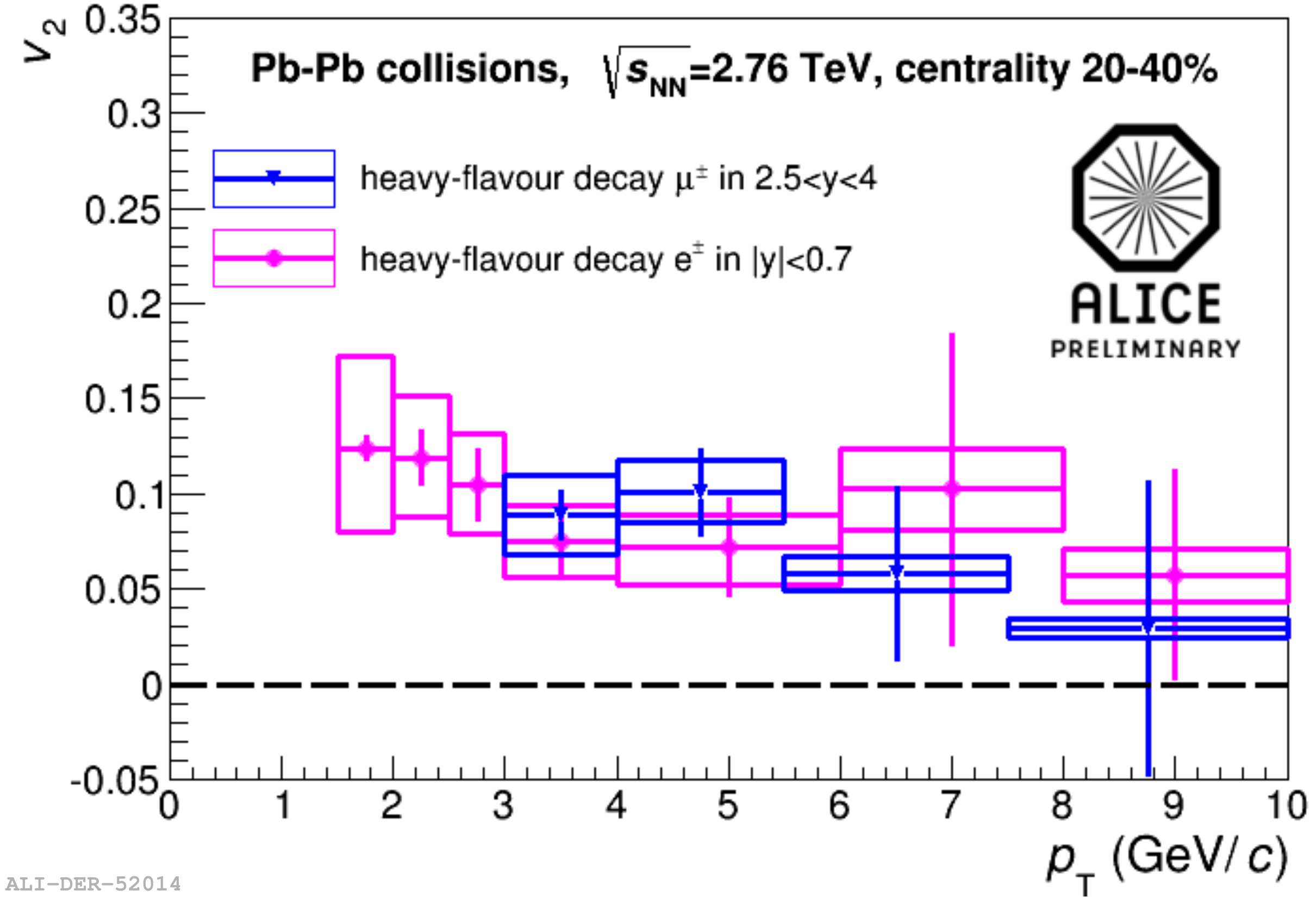}

\end{subfigure}
\caption{(Color online) Left: \pt~ integrated \v of heavy-flavour decay muons as a function of collision centrality measured in Pb--Pb collisions at \sqsnn = 2.76 TeV in the \pt~ range 3--10 GeV/$c$ in 2.5 $ < y < $ 4. Right: Comparison of the measured \v as a function of \pt~ in Pb--Pb collisions in the centrality class 20--40$\%$ of the muons (circles) measured in the forward region (2.5 $ < y < $ 4) and electrons (triangles) measured at central rapidity ($|y| < $ 0.7).}
\label{fig:leptonv2}       
\end{figure} 

In the case of a non-central heavy-ion collision, an azimuthal anisotropy in momentum space can be observed in the measured final state particle distributions. The particle distribution can be expanded in a Fourier series \cite{voloshin}
\begin{equation}
E \frac{\mathrm{d}^3N}{\mathrm{d}p^3} = \frac{1}{2\pi} \frac{\mathrm{d}^2N}{p_\mathrm{T} \mathrm{d}p_\mathrm{T} \mathrm{d}y} \left(1 + 2 \sum_{n=1}^{\infty} \nu_n \cos [n(\varphi - \Psi_{RP})] \right)
\end{equation}
\noindent
Here $\Psi_{RP}$ is the reaction plane azimuthal angle and the second coefficient, $\nu_2$, is the elliptic flow. The reaction plane is defined by the impact parameter and beam directions.

Fig. \ref{fig:leptonv2} (left) shows the \pt~ integrated (3 $<$ \pt~ $< $ 10 GeV/$c$) heavy-flavour decay muon \v as a function of centrality. \v increases from central (0--10$\%$) to semi-central (20--40$\%$) collisions, with a positive \v observed in semi-central collisions. The \v for the most central collisions approached zero and is consistent with the expectation that the initial geometrical anisotropy vanishes with increased centrality. Shown in Fig. \ref{fig:leptonv2} (right) is the comparison of heavy-flavour decay electron \v as a function of \pt~ measured in $|y| < $ 0.7 for the centrality class 20--40$\%$, to muons measured in the same centrality class but in the range 2.5 $ < y < $ 4. The mid-rapidity \v of electrons from heavy-flavour hadron decays and the muons from such decays in the forward region are compatible within the experimental uncertainties.

\begin{figure}
\includegraphics[width=1\linewidth]{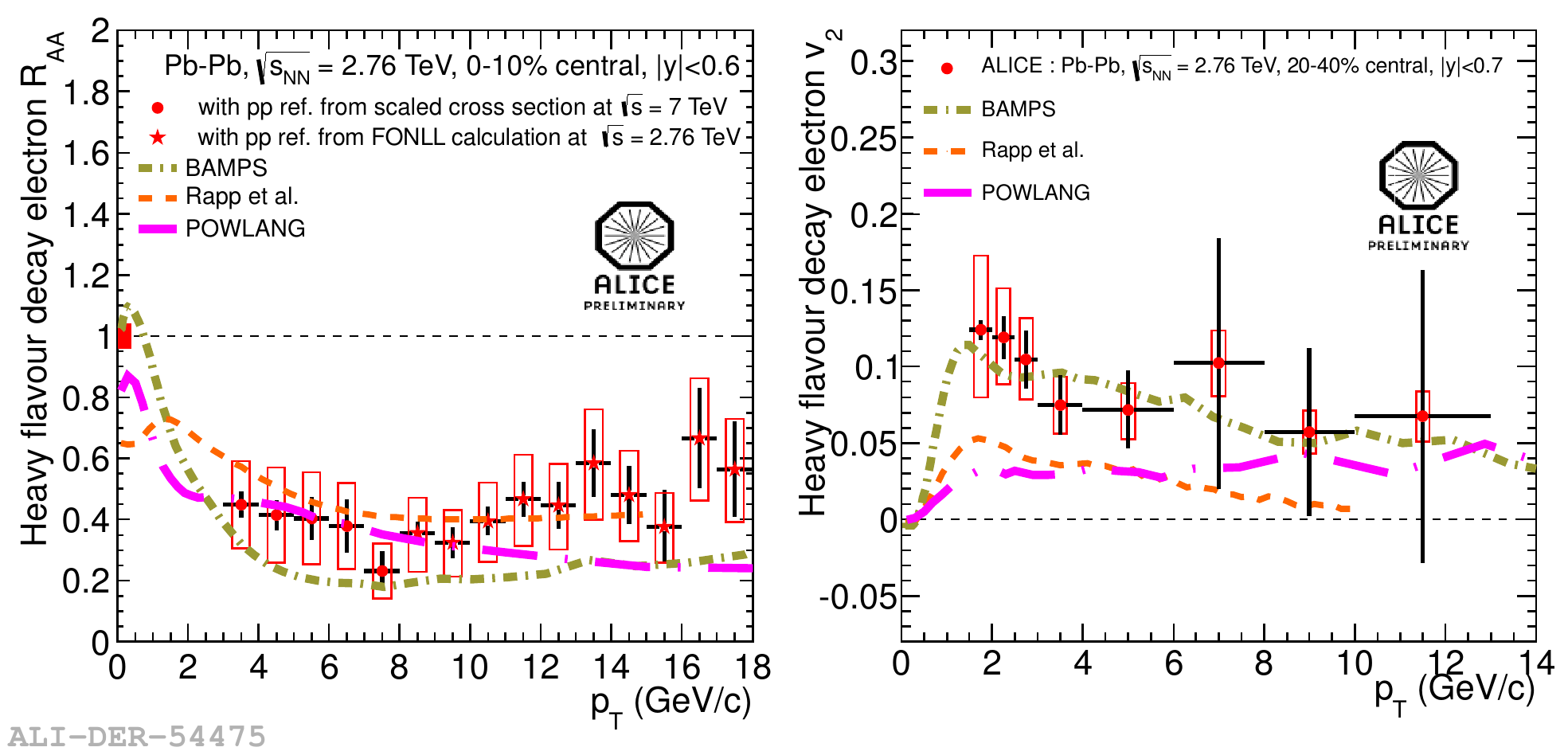}
\caption{(Color online) Left: Heavy-flavour decay electron \raa as a function of \pt~ measured in the 0--10$\%$ centrality class in  $|y| < $ 0.6. The circles are where the pp reference is taken as the scaled spectrum at \sqs = 7 TeV, while for the star symbols the reference is a FONLL calculation at \sqs = 2.76 TeV. The results are compared to three parton transport models: BAMPS \cite{bamps}, Rapp et al. \cite{rapp}, and POWLANG \cite{powlang}. Right: Heavy-flavour electron \v as a function of \pt~ measured in the 20-40$\%$ centrality class in $|y| < $ 0.6. The results are compared to the aforementioned models.}
\label{fig:compare}       
\end{figure}

In Fig. \ref{fig:compare} the measured electron R$_{\mathrm{AA}}$ (left)  and $\nu_2$ (right) are compared to three parton transport models: BAMPS, which includes collisional in-medium energy loss and mimics the radiative processes by increasing the elastic cross section \cite{bamps}, Rapp et al., which incorporates energy loss using collisional processes via a non-perturbative T-matrix approach \cite{rapp}, and POWLANG, which is based on the Langevin equation with collisional energy loss in-medium \cite{powlang}. The comparison of the individual models reveals the difficulty they have in describing both observables simultaneously and has prompted refinement of the models, e.g. Rapp et al. \cite{rapp2014}.


%
%

\section{Measurements in p--Pb collisions}
\label{pPb}

\begin{figure}
\begin{subfigure}[b]{.5\textwidth}
\includegraphics[width=1\linewidth]{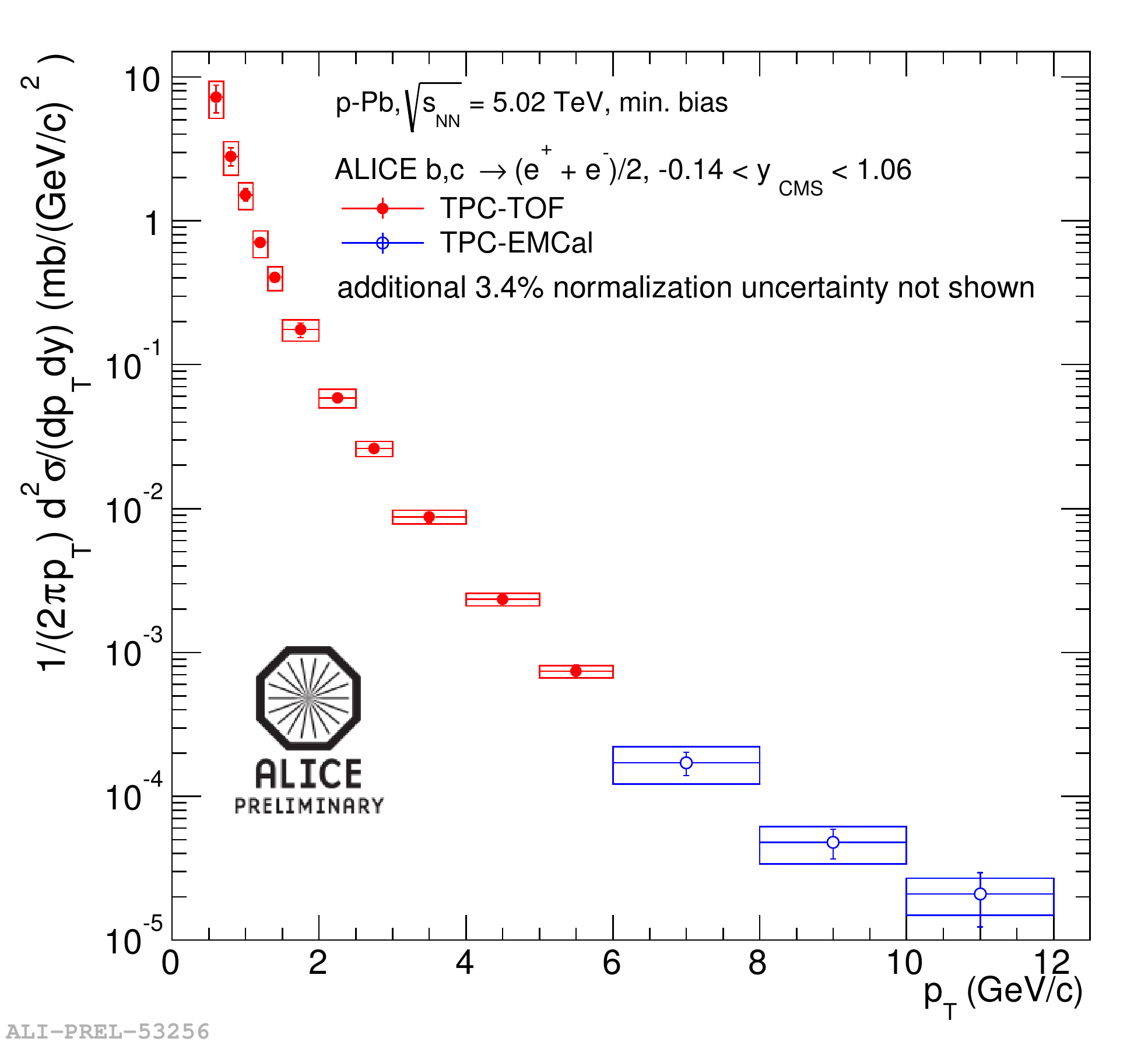}
\end{subfigure}%
\begin{subfigure}[b]{.5\textwidth}
  \includegraphics[width=1\linewidth]{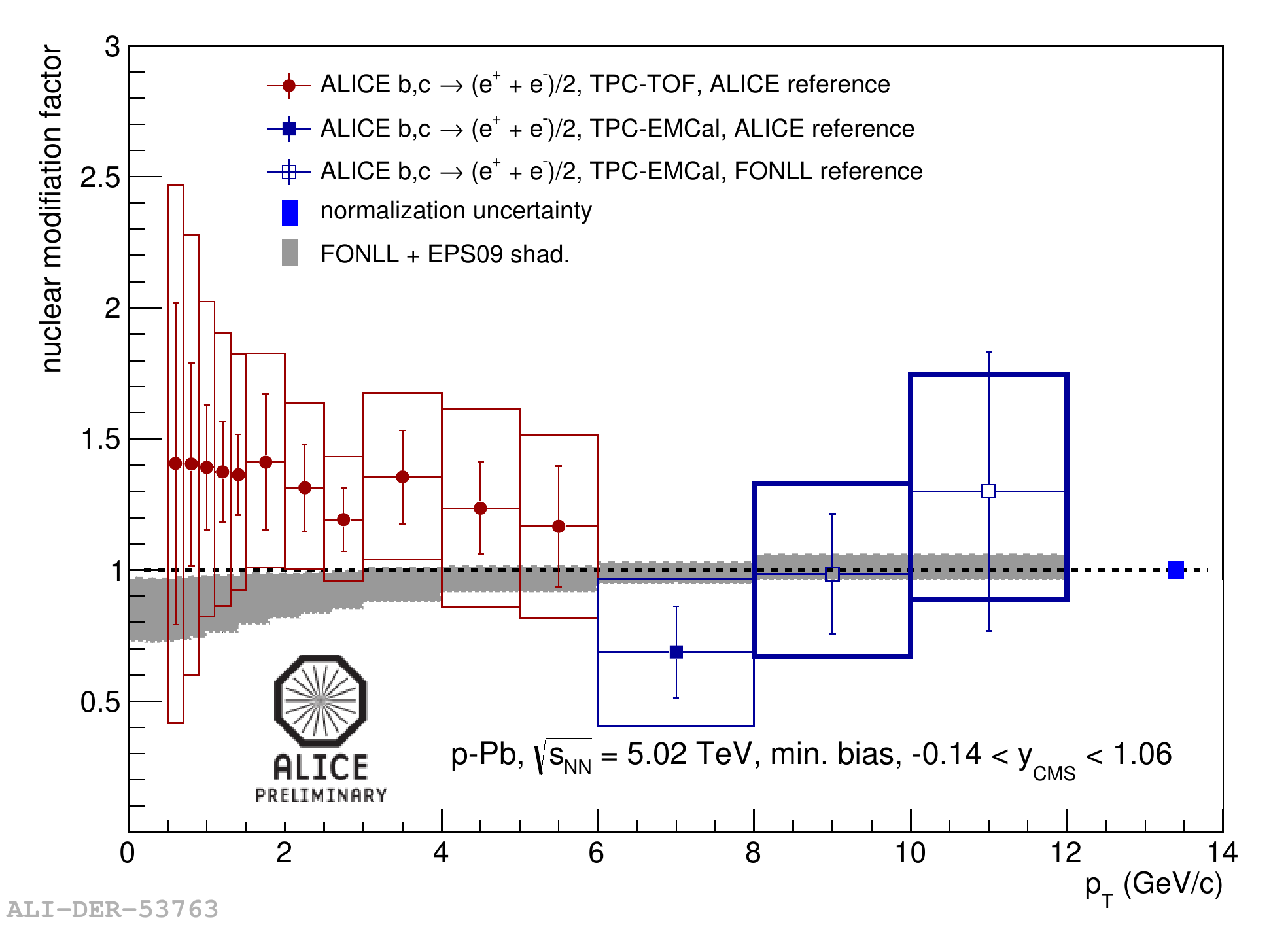}
\end{subfigure}
\caption{(Color online) Left: \pt-differential production cross section of electrons from heavy-flavour hadron decays in minimum bias p--Pb collisions at \sqsnn = 5.02 in the range $|\eta| < $ 0.6. TeV. Right: R$_{\mathrm{pPb}}$ of electrons from the decay of heavy-flavour hadrons as a function of \pt. The solid band is the prediction of FONLL including the EPS09 paramatrization of nuclear modification of PDFs. For both figures, the circle symbols represent the electrons identified using TPC-TOF, while those identified using TPC-EMCal are shown as square symbols.}
\label{fig:elecRpPb}       
\end{figure}

Heavy-flavour decay electrons were measured in minimum bias p--Pb collisions at \sqsnn = 5.02 TeV in the rapidity interval $|\eta| < $ 0.6. Using the TPC and TOF for electron identification the measurement spans the \pt~ range 0.5--6 GeV/$c$, while electrons identified utilizing the TPC and EMCal were measured in the \pt~ range 2--12 GeV/$c$. The overlap region of these results is in agreement with the TPC--TOF based measurement featuring smaller statistical uncertainties. In Fig. \ref{fig:elecRpPb} the resulting \pt-differential cross section is shown.

The nuclear modification factor R$_{\mathrm{pPb}}$ is shown in Fig. \ref{fig:elecRpPb} (right). For \pt~ $< $ 8 GeV/$c$ the p--Pb spectrum is compared to the pp reference spectrum measured at \sqs = 7 TeV and scaled to 5.02 TeV, while above 8 GeV/$c$ the FONLL calculation was used as the reference. Within the large experimental uncertainties, the result can be interpreted as in agreement with nuclear modification of the parton distribution functions predicted by EPS09 \cite{eps}. This result points to small cold nuclear matter effects and indicates that the suppression observed in Pb--Pb is a hot medium effect.

\section{Conclusions}
\label{con}

The present status of heavy-flavour studies via semi-leptonic decay channels in ALICE has been shown. In Pb--Pb collisions the nuclear modification factor exhibits a strong suppression in the most central collisions, which indicates significant energy loss of heavy quarks in the produced medium. \v measurements of heavy-flavour decay leptons in Pb--Pb collisions indicate a positive $\nu_2$ in semi-central collisions, suggesting that heavy quarks experience the collective motion of the QCD medium. Finally, R$_{\mathrm{pPb}}$ is consistent with small cold nuclear matter effects.

\bibliography{HFleptons}
\bibliographystyle{elsarticle-num.bst}

\end{document}